\documentclass[twocolumn,epjc3]{svjour3}  
\smartqed
\usepackage[utf8,latin1]{inputenc}
\newcommand{\qm}[1]{``#1''}
\usepackage{amsmath}
\usepackage{amssymb}
\usepackage{mathabx}
\usepackage{mathrsfs}  
\usepackage{cases}
\usepackage{revsymb}
\usepackage{tabularx}
\usepackage{colortbl}
\usepackage{graphicx}
\usepackage{dcolumn}
\usepackage{xcolor}
\usepackage{bm}
\usepackage{hyperref}
\hypersetup{colorlinks, linkcolor={red},citecolor={blue},urlcolor={blue}}  
\usepackage[square,numbers,sort&compress]{natbib}
\usepackage{hyperref}

  \def\nn{\nonumber} 
\newcommand{\dd}{{\rm d}}
\newcommand{\OO}{{\rm O}}

\RequirePackage{graphicx}
\journalname{Eur. Phys. J. C}

\begin{document}

\title{First post-Newtonian $N$-body problem in Einstein-Cartan theory with the Weyssenhoff fluid: Lagrangian and first integrals}
\titlerunning{First post-Newtonian $N$-body problem in Einstein-Cartan theory with the Weyssenhoff fluid: Lagrangian and first integrals}

\author{Emmanuele Battista\thanksref{e1,e2,addr1}
\and
Vittorio De Falco\thanksref{e3,addr2,addr3}
\and
Davide Usseglio\thanksref{e4,addr2,addr3}}

\thankstext{e1}{e-mail: emmanuele.battista@univie.ac.at}
\thankstext{e2}{e-mail: emmanuelebattista@gmail.com}
\thankstext{e3}{e-mail: vittorio.defalco-ssm@unina.it}
\thankstext{e4}{e-mail: davide.usseglio-ssm@unina.it}

\authorrunning{Battista, De Falco, Usseglio (2022)}

\institute{Department of Physics, University of Vienna, Boltzmanngasse 5, A-1090 Vienna, Austria \label{addr1}
\and
Scuola Superiore Meridionale, Largo San Marcellino 10, 80138 Napoli, Italy\label{addr2}
\and
Istituto Nazionale di Fisica Nucleare, Sezione di Napoli, Complesso Universitario di Monte S. Angelo, Via Cintia Edificio 6, 80126 Napoli, Italy \label{addr3}}

\date{Received: \today / Accepted: }

\maketitle

\begin{abstract}
The rotational dynamics of an $N$-body system at the first post-Newtonian order in Einstein-Cartan theory
is derived. This result is achieved by performing the point-particle limit of the  equations of motion of the  Weyssenhoff fluid, which models the quantum spin effects residing inside the bodies. For the special case of  binary systems, we determine the Lagrangian function and the resulting first integrals underlying the translational dynamics and the spin precession. 
\end{abstract}

\section{Introduction}
\label{sec:intro}
The Lagrangian formalism is widely exploited in physics, because it is able to capture  all the  dynamical features of the system under study. In the context of classical (non-dissipative) mechanics, this approach entails the following main advantages \cite{Goldstein2002}: ($i$) the Lagrangian function deals with the energy of the system, instead of the forces acting upon it; ($ii$) given the Lagrangian, assigned the forces, and chosen the generalized coordinates, one can directly characterize the dynamics via the Euler-Lagrange equations; ($iii$) the symmetries of the Lagrangian  can be associated with the existence of  first integrals thanks to   the Noether theorem.  

The nature of the Euler-Lagrange equations changes depending on the context to be investigated. In  the case of the \emph{direct problem} (where they are computed through the given Lagrangian), these are second-order ordinary differential equations. On the other hand, for the \emph{inverse problem} (given the dynamics, the Lagrangian must be determined), they become a set of second-order partial differential equations \cite{Santilli1978,Santilli1979,Lopuszanski1999}. 

In General Relativity (GR), the inverse approach is widely used to determine the Lagrangian associated to the motion of an $N$-body system. 
Due to the non-linear hyperbolic structure of GR, the aforementioned problem cannot be solved analytically, since it yields retarded-partial-integro differential equations \cite{Maggiore:GWs_Vol1,Blanchet2014,Poisson-Will2014}. However, these mathematical complications can be circumvented by resorting to solid and well-founded approximation schemes \cite{Maggiore:GWs_Vol1,Blanchet2014}. First of all, the gravitational source, which is first modelled 
as a continuous smooth hydrodynamical distribution of matter, is assumed to be post-Newtonian (PN), namely it is slowly moving, weakly self-gravitating, and weakly stressed. Thus, its dynamics can be studied by employing the \emph{PN approximation method} in the so-called near zone. Furthermore, by supposing that the source consists of $N$  mutually well separated fluid bodies, the point-particle limit can be invoked. This pattern may find different applications both in astrophysics and cosmology. In particular, it represents a useful mean to analyse the dynamics of inspiralling compact binaries, which represent the main astrophysical sources of gravitational waves (GWs).

In 1917, Lorentz and Droste determined for the very first time the Lagrangian and the equations of motion for the GR two-body problem at the first post-Newtonian (1PN) order \cite{Droste1917,Lorentz1937}. In  1938, Einstein, Infeld, and Hoffmann (EIH) \cite{Einstein1938,Infeld1960} re-derived these results for $N$ bodies by employing the surface integral method. At  2PN level, some subtleties come into play. Indeed,  the 2PN-accurate GR dynamics of an $N$-body system in harmonic coordinates can be derived from a \emph{generalized Lagrangian}, which, apart from the positions and velocities, depends also on the accelerations of the particles \citep{Damour1983}. This result has been established, on general grounds, in  Ref.  \citep{Martin1979}, where it has been rigourously proved that, under certain hypotheses, a system of $N$ non-spinning objects interacting via gravitational  forces cannot be examined through the usual Lagrangian picture if  2PN corrections occur in the equations of motion. In the case of spinning interacting bodies, the (spin-orbit) Lagrangian depends on the accelerations already at 1PN level (see e.g., Ref. \citep{Kidder1993} and references therein). 

In this article, we investigate  the $N$-body problem at 1PN level in  Einstein-Cartan (EC) theory, which configures as an extension of GR, where the non-Riemannian part of the connection is related to the microscopic spin density of the source \cite{Hehl1976_fundations}. Hereafter, the term \qm{spin} will refer to the quantum intrinsic angular momentum of bodies. 

This work represents an important advancement in our research program aimed at studying GWs and related phenomena  in EC theory. In fact, we have first studied the 1PN GW generation problem by means of the Blanchet-Damour formalism in Ref. \cite{Paper1}. Then, we have considered an explicit application by modelling the gravitational source and the underlying spin effects through the  Weyssenhoff fluid \cite{Paper2}. By means of the point-particle procedure, we have also provided the 1PN-accurate formulas of the source and the radiative multipole moments characterizing an $N$-body system. The study of the 1PN dynamics has been started out in Ref. \cite{Paper3}, where we have determined the equations governing the translational motion of $N$ objects subject to their mutual gravitational attraction. In this paper, we complete this analysis.  Firstly, we derive the $N$-body rotational equations of motion in EC theory at the 1PN level (see Sect. \ref{sec:N-body-rotational}). Subsequently, we reconstruct the Lagrangian and calculate the  first integrals governing the  dynamics of a spinning binary system at 1PN order (see Sect. \ref{sec:Lagrangian-EC-theory}). Last, we give a  summary of our findings and  present   future perspectives of our work (see Sect. \ref{sec:end}).

\emph{Notations.} We use metric signature  $(-,+,+,+)$. Greek indices take values  $0,1,2,3$, while lowercase Latin ones $1,2,3$. The determinant of the metric $g_{\mu \nu}$ is denoted by $g$. $\varepsilon_{kli}$ is the total antisymmetric Levi-Civita symbol. The spacetime coordinates are $x^\mu = (ct,\boldsymbol{x})$. Four-vectors are written as $a^\mu = (a^0,\boldsymbol{a})$, and $\boldsymbol{a} \cdot \boldsymbol{b}:= \delta_{lk}a^l b^k$, $\vert \boldsymbol{a} \vert\equiv a :=  \left(\boldsymbol{a} \cdot \boldsymbol{a}\right)^{1/2}$, and $\left(\boldsymbol{a} \times \boldsymbol{b}\right)^i := \varepsilon_{ilk} a^l b^k$. The symmetric-trace-free projection of a tensor $A^{ij\dots k}$ is indicated with $A^{\langle ij\dots k \rangle }$. Round (respectively, square) brackets around a pair of indices stands for the usual symmetrization (respectively, antisymmetrization) procedure, i.e., $A_{(ij)}=\frac{1}{2}(A_{ij}+A_{ji})$ (respectively, $A_{[ij]}=\frac{1}{2}(A_{ij}-A_{ji})$). A over-hat symbol refers to quantities framed in GR. The $N$ bodies and all the related quantities are labelled with capital Latin indices, such as $A,B,C=1, \dots, N$.

\section{$N$-body problem} 
\label{sec:N-body-rotational}
EC gravity model represents   the gauge theory of the Poincar\'e group, the semidirect product of the translation and   the Lorentz groups. In this  framework, the sources of the gravitational field  are represented by both the energy-momentum tensor (i.e., the translational Noether current) and the spin angular-momentum tensor (i.e., the Noether intrinsic  rotational current). \emph{A distinguished feature of EC pattern is the prediction of a spin-spin contact interaction of gravitational
origin} \cite{Hehl1976_fundations}. 

EC  theory is defined on a  Riemann-Cartan spacetime endowed with a symmetric metric tensor $g_{\alpha \beta}$ and the most general metric-compatible affine connection $\Gamma^\lambda_{\mu \nu}:=\hat{\Gamma}^{\lambda}_{\mu \nu}-K_{\mu \nu}^{\ \ \ \lambda}$, where $\hat{\Gamma}^{\lambda}_{\mu \nu}=\hat{\Gamma}^{\lambda}_{(\mu \nu)}$ is the \emph{Levi-Civita connection} and $K_{\mu \nu}^{\ \ \ \lambda}$ the \emph{contortion tensor}. The antisymmetric part of the connection $\Gamma^\lambda_{[\mu \nu]}:= S_{\mu \nu}^{\ \ \ \lambda}$ defines the so-called \emph{Cartan torsion tensor}. 

In this section, we deal with the $N$-body problem in EC theory at 1PN order by considering the Weyssenhoff fluid as the  model of the spinning matter. After having reviewed the translational equations in Sect. \ref{sec:EoMs_1PN},  we tackle the point-particle limit of the rotational motion   in Sect. \ref{Sec:rotational-motion}. Last, a physical discussion concerning the structure integrals occurring in the $N$-body dynamics is provided in Sect. \ref{Sec:inner-structure-dependent quantities}.

\subsection{The translational motion}
\label{sec:EoMs_1PN}
The semiclassical description of a spinning perfect fluid within the EC theory can be obtained by means of the Weyssenhoff model  \cite{Obukhov1987,Boehmer2006}. In this approach, the fluid is  characterized by the spin angular momentum tensor
\begin{align}
\tau_{\alpha\beta}{}^\gamma&=s_{\alpha\beta}u^\gamma,
\label{eq:spin-tensor-fluid}
\end{align}
and satisfies the \emph{Frenkel condition} 
\begin{equation} \label{eq:Frenkel_condition}
\tau_{\alpha\beta}{}^\beta= s_{\alpha\beta}\,u^\beta=0,
\end{equation}
where $s_{\alpha\beta}=s_{[\alpha\beta]}$ and $ u^\alpha = \frac{u^0}{c} \left(c,\boldsymbol{v}\right)$ (with $\boldsymbol{v} := {\rm d}\boldsymbol{x}/{\rm d}t$ the coordinate velocity) denote the spin density tensor  and the timelike four-velocity vector of the fluid, respectively. We note that Eq. \eqref{eq:Frenkel_condition} amounts to require that the torsion tensor has  vanishing trace (i.e., $S^{\alpha\mu}{}_\mu=0$), a condition which fulfils a crucial role in our analysis  (see Refs. \cite{Paper1,Paper2}, for further details). 

In order to work out the 1PN translational motion of the $N$-body system, we need to apply the point-particle limit to the (continuous)  equations ruling the translational dynamics of the  Weyssenhoff fluid \cite{Paper2,Paper3}. We will employ this procedure by supposing that the $N$ objects composing the system are: (1) reflection symmetric about their center of mass; (2) in stationary equilibrium; (3) mutually well separated. 

Let us adopt the following definitions:
\begin{subequations}
\begin{align}
m_A &:= \int_A  \dd^3 \boldsymbol{x} \; \rho^\star,
\label{eq:material-mass-A}
\\
\varepsilon_{jki}\, {}^{(n)}s_A^i(t) &:=\int_A {\rm d}^3 \boldsymbol{x} \, {}^{(n)}s_{jk}, \qquad (n=1,3),
\label{eq:spin-vector-body-A}
\\
\boldsymbol{x}_A(t)&:=\dfrac{1}{m_A} \int_A \dd^3 \boldsymbol{x} \;  \rho^\star \boldsymbol{x}, 
\\
\boldsymbol{v}_A(t)&:=\dfrac{\dd \boldsymbol{x}_A}{\dd t}=\dfrac{1}{m_A} \int_A \dd^3   \boldsymbol{x} \;  \rho^\star \boldsymbol{v}, 
\\
\boldsymbol{a}_A(t)&:=\dfrac{\dd \boldsymbol{v}_A}{\dd t}=\dfrac{1}{m_A} \int_A \dd^3 \boldsymbol{x} \; \rho^\star \frac{\dd \boldsymbol{v}}{\dd t},
\end{align}
\end{subequations}
which represent the (conserved) material mass, the spin vector (with ${}^{(n)}s_{ij}= {\rm O}\left(c^{1-n}\right)={}^{(n)}s^{j}$ \cite{Paper1,Paper2}), the center of mass, the center of mass velocity, and the center of mass acceleration of the body $A$, respectively. In the above formulas, $\rho^\star := \frac{u^0}{c} \sqrt{-g} \rho = \rho + \OO\left(c^{-2}\right)$ is the coordinate rest-mass density of the fluid expressed in terms of rest-mass density $\rho$. 

Bearing in mind the above equations, the   harmonic-coordinate translational dynamics of the system is encoded in  the following expression \cite{Paper3}: 
\begin{align} 
a_A^i &=a_{A,{\rm EIH}}^i + \frac{4}{c^2}\sum_{B \neq A} \frac{G}{r_{AB}^3} \Biggl\{ 2 \Bigl[ \left(\boldsymbol{v}_B-\boldsymbol{v}_A\right) \times \boldsymbol{s}_B\Bigr]^i  
\nn \\
&+3 n_{AB}^i \; \boldsymbol{s}_B \cdot  \left[\boldsymbol{n}_{AB} \times \left(\boldsymbol{v}_A-\boldsymbol{v}_B \right) \right]
\nn \\
&+3\left( \boldsymbol{n}_{AB} \times \boldsymbol{s}_B\right)^i \left(\boldsymbol{v}_A - \boldsymbol{v}_B\right) \cdot \boldsymbol{n}_{AB} \Biggr\}
\nn \\
&-\frac{6}{c^2} \sum_{B \neq A} \frac{G M_B}{M_Ar_{AB}^3} \Biggl\{  \Bigl[ \left(\boldsymbol{v}_A-\boldsymbol{v}_B\right) \times \boldsymbol{s}_A\Bigr]^i  
\nn \\
&-2 n_{AB}^i \; \boldsymbol{s}_A \cdot  \left[\boldsymbol{n}_{AB} \times \left(\boldsymbol{v}_A-\boldsymbol{v}_B \right) \right]
\nn \\
&+\left( \boldsymbol{n}_{AB} \times \boldsymbol{s}_A\right)^i \left(\boldsymbol{v}_B - \boldsymbol{v}_A\right) \cdot \boldsymbol{n}_{AB} \Biggr\}
\nn \\
&-\frac{12}{c^2} \sum_{B \neq A} \frac{G }{M_Ar_{AB}^4} \Biggl \{ s^i_A \left(\boldsymbol{n}_{AB}\cdot \boldsymbol{s}_B\right)+s^i_B \left(\boldsymbol{n}_{AB}\cdot \boldsymbol{s}_A\right)
\nn \\
&+ n_{AB}^i \Bigl[ \boldsymbol{s}_A \cdot \boldsymbol{s}_B - 5 \left(\boldsymbol{n}_{AB}\cdot \boldsymbol{s}_A\right)\left(\boldsymbol{n}_{AB}\cdot \boldsymbol{s}_B\right)\Bigr] \Biggr\}
\nn \\
&+ \OO\left(c^{-4}\right),
\label{eq:EC-body-A-equation-of-motion-2}
\end{align}
where $a_{A,{\rm EIH}}^i$ is the EIH acceleration of the object $A$ (see Appendix \ref{sec:GR-Appendix}, for further details) and $M_A=m_A +{\rm O}\left(c^{-2}\right)$ its (conserved) total mass-energy; moreover, we have taken into account that the spin vector admits the PN structure  
 \begin{align}
\label{eq:spin_vector_PN_expansion}
\boldsymbol{s}_A = {}^{(1)}\boldsymbol{s}_A +{\rm O}(c^{-2}),
\end{align}
and we have introduced the following variables:
\begin{align}
\boldsymbol{r}_{AB} &:= \boldsymbol{x}_A - \boldsymbol{x}_B, \qquad  \boldsymbol{n}_{AB} := \dfrac{\boldsymbol{r}_{AB}}{r_{AB}}.
\end{align} 
Equation \eqref{eq:EC-body-A-equation-of-motion-2}, jointly with the conservation law   $\dd \boldsymbol{s}_A / \dd t = {\rm O}\left(c^{-2}\right)$, completely determines the dynamics of the $N$-body system at 1PN level. As set out in Ref. \cite{Paper3}, the lack of  contributions due to the inner details of the bodies  can be interpreted as a hint for the validity of the \emph{effacing  principle} at 1PN order.

\subsection{The rotational motion}\label{Sec:rotational-motion}
The rotational dynamics of the Weyssenhoff fluid in EC theory is ruled by the exact equation \cite{Paper2}
\begin{align} \label{eq:rotational_fluid_equation}
\hat{\nabla}_\lambda \left(s_{\mu \nu} u^\lambda\right)  & = \dfrac{a^\sigma}{c^2} \left(u_\mu   s_{ \sigma \nu}- u_\nu   s_{ \sigma \mu} \right),
\end{align}
where $a^\sigma$ is the fluid acceleration. If  we exploit  (the PN expansion of) the Frenkel condition \eqref{eq:Frenkel_condition} and the PN series of the spin density 
\begin{align}
\label{eq:spin_density_PN_expansion}
s_{ij} &= {}^{(1)}s_{ij} + {}^{(3)}s_{ij} +{\rm O}(c^{-4}),
\end{align}
Eq. \eqref{eq:rotational_fluid_equation} yields at 1PN level  and   in harmonic coordinates
\begin{align}
\dfrac{{\rm d}}{{\rm d}t} & s_{ij} +s_{ij}\partial_k v^k +\frac{1}{c^2} \Biggl [ s_{ij} \partial_t \hat{\mathscr{U}} + 2  s_{ij} v^k \partial_k \hat{\mathscr{U}}
\nonumber \\
&-  \dfrac{\partial_k P}{\rho^\star} \left( v^j s_{ki}-v^i s_{kj} + v^k s_{ij} \right) + 2 s_{ki} \left(\partial_k \hat{\mathscr{U}}_j \right.
\nonumber \\
&\left.+\partial_k \Sigma_j -\partial_j \hat{\mathscr{U}}_k -\partial_j \Sigma_k+ v^k \partial_j \hat{\mathscr{U}} -\frac{v^j}{2} \partial_k \hat{\mathscr{U}}\right)
 \nonumber \\
 &-2 s_{kj} \left(\partial_k \hat{\mathscr{U}}_i + \partial_k \Sigma_i -\partial_i \hat{\mathscr{U}}_k -\partial_i \Sigma_k+ v^k \partial_i \hat{\mathscr{U}} \right. 
 \nonumber \\
& \left. -\frac{v^i}{2} \partial_k \hat{\mathscr{U}}\right) \Biggr]  ={\rm O}\left(c^{-4}\right),
 \label{eq:1PN-spin-equation-potential-explicit}
\end{align}
where $P$ is the fluid  pressure, and we have exploited the 0PN-accurate  equations  ${\rm d} v^k/ {\rm d}t = \partial_k \hat{\mathscr{U}} -(1/\rho^\star) \partial_k P + {\rm O}\left(c^{-2}\right)$ and  ${\rm d}s_{ij}/ {\rm d}t + s_{ij} \partial_k v^k={\rm O}\left(c^{-2}\right)$,  along with the definition of the potentials 
\begin{subequations}
\label{eq:potentials-U-U-i-Sigma-i}
\begin{align}
\hat{\mathscr{U}}\left(t,\boldsymbol{x}\right) &:=G  \int \dfrac{{\rm d}^3\boldsymbol{x}^\prime}{|\boldsymbol{x}-\boldsymbol{x}^\prime|}\rho^{\star \prime},
\label{eq:curly-U-potential-EC-theory}
\\
\hat{\mathscr{U}}_i \left(t,\boldsymbol{x}\right) &:=G  \int \dfrac{{\rm d}^3\boldsymbol{x}^\prime}{|\boldsymbol{x}-\boldsymbol{x}^\prime|}\rho^{\star \prime} v^{\prime i},
\label{eq:curly-U-i-potential-EC-theory}
\\
\Sigma_i \left(t,\boldsymbol{x}\right) &:= G \int \dd^3 \boldsymbol{x}^\prime \dfrac{(x - x^\prime)_k}{|\boldsymbol{x}-\boldsymbol{x}^\prime|^3} \,s^\prime_{ki},
\label{eq:Sigma-i-potential-EC}
\end{align}
\end{subequations}
the primed variables being evaluated at time $t$ and position $\boldsymbol{x}^\prime$. 

At this stage, we can apply the point-particle procedure to Eq. \eqref{eq:1PN-spin-equation-potential-explicit}. This process relies mainly on: (i) the separation  of the potentials into internal and external components; (ii) the analysis of the contributions introduced by the internal potentials $\hat{\mathscr{U}}_A$, $\hat{\mathscr{U}}_{i,A}$, and $\Sigma_{i,A}$, which lead, in general, to structure-dependent integrals; (iii) the evaluation of the derivatives of the external potentials $\hat{\mathscr{U}}_{\neg A}$, $\hat{\mathscr{U}}_{i,\neg A}$, and $\Sigma_{i,\neg A}$ in $\boldsymbol{x}=\boldsymbol{x}_A$ (see Ref. \cite{Paper3}, for further details). All the computations are performed by exploiting the aforementioned  hypotheses (1)-(3) (see Sect. \ref{sec:EoMs_1PN}), where, in particular, hypothesis (3) permits to neglect terms of fractional order $(\ell_A/d_A)^2$ or $(\ell_A/r_{AB})^2$, where $\ell_A$ denotes the typical linear  dimension of  $A$ and and $d_A := \vert \boldsymbol{x} - \boldsymbol{x}_A \vert$. In this way, after a lengthy calculation we obtain the following expression: 
\begin{align}
\varepsilon_{ijl}  \dfrac{{\rm d}}{{\rm d}t} & \left[s_A^l + \frac{s_A^l}{c^2}\left(\hat{\mathscr{U}}_{\neg A} +\frac{ v_A^2}{2}\right)\right]   +\frac{1}{c^2} \Biggl [ 2 \varepsilon_{kil} s^l_A \left(\partial_k \hat{\mathscr{U}}_{j,\neg A} \right.
\nonumber \\
&\left. -\partial_j \hat{\mathscr{U}}_{k,\neg A} +\partial_k \Sigma_{j,\neg A} -\partial_j \Sigma_{k,\neg A}+ v_A^k \partial_j \hat{\mathscr{U}}_{\neg A} \right.
\nonumber \\
&\left. -\frac{v_A^j}{2} \partial_k \hat{\mathscr{U}}_{\neg A}\right) + \mathcal{Q}_A^{ij}-\mathcal{Z}_A^{ij} - 2 \varepsilon_{kjl} s^l_A \left(\partial_k \hat{\mathscr{U}}_{i,\neg A} \right.
\nonumber \\
&\left. -\partial_i \hat{\mathscr{U}}_{k,\neg A} +\partial_k \Sigma_{i,\neg A} -\partial_i \Sigma_{k,\neg A}+ v_A^k \partial_i \hat{\mathscr{U}}_{\neg A} \right.
\nonumber \\
&\left. -\frac{v_A^i}{2} \partial_k \hat{\mathscr{U}}_{\neg A}\right) - \mathcal{Q}_A^{ji}+\mathcal{Z}_A^{ji} \Biggr] ={\rm O}\left(c^{-4}\right),
 \label{eq:PPL-spin-equation-1}
\end{align}
where $\mathcal{Q}_A^{ij}=\mathcal{Q}_A^{ji}$ and $\mathcal{Z}_A^{ij}\neq\mathcal{Z}_A^{ji}$ are structure integrals, which will be discussed in Sect. \ref{Sec:inner-structure-dependent quantities}. It is important to stress that, in the above equation, the spin vector has the PN form 
\begin{align}
\boldsymbol{s}_A &= {}^{(1)}\boldsymbol{s}_A + {}^{(3)}\boldsymbol{s}_A +{\rm O}(c^{-4}),
\label{eq:PN-vector-s-A}
\end{align}
and  all the external potentials are evaluated at $\boldsymbol{x}=\boldsymbol{x}_A$. If we multiply Eq. \eqref{eq:PPL-spin-equation-1} by $\varepsilon_{ijp}$ and take into account the formulas for the   derivatives of the external potentials (see Sect. 2.2.4 in Ref. \cite{Paper3}), we obtain the 1PN-accurate spin precession equation 
\begin{align}
\frac{\dd }{\dd t} & \left\{\boldsymbol{s}_A + \frac{1}{c^2} \left[\boldsymbol{s}_A \hat{\mathscr{U}}_{\neg A} +\frac{1}{2} \left(\boldsymbol{s}_A \cdot \boldsymbol{v}_A\right)\boldsymbol{v}_A \right]\right\} 
\nonumber  \\
&= \boldsymbol{\Omega}_A \times \boldsymbol{s}_A+ \OO\left(c^{-4}\right), 
\label{eq:spin-precession-1}
\end{align}
with
\begin{align}
 \boldsymbol{\Omega}_A := \boldsymbol{\Omega}_A^{\rm SO} + \boldsymbol{\Omega}_A^{\rm SS}. 
\end{align}
The spin-orbit  (SO) piece stems from the derivatives of $\hat{\mathscr{U}}_{\neg A}$ and  $\hat{\mathscr{U}}_{i,\neg A}$ and reads as
\begin{align}
\boldsymbol{\Omega}_A^{\rm SO} &= \frac{1}{2c^2} \sum_{B \neq A} \frac{G M_B}{r_{AB}^2} \left[\boldsymbol{n}_{AB} \times \left(3\boldsymbol{v}_A-4\boldsymbol{v}_B\right)\right],
\end{align}
while the spin-spin (SS) contribution comes from the derivatives of   $\Sigma_{i,\neg A}$ and is given by
\begin{align}
\boldsymbol{\Omega}_A^{\rm SS} &= \frac{1}{c^2} \sum_{B \neq A} \frac{2G }{r_{AB}^3} \left[ 3 \left(\boldsymbol{n}_{AB} \cdot \boldsymbol{s}_B\right)\boldsymbol{n}_{AB} -\boldsymbol{s}_B \right].
\label{eq:Omega-SS}    
\end{align}

Similarly to the GR framework \cite{Poisson-Will2014}, we can define the refined spin vector 
\begin{align}
\bar{\boldsymbol{s}}_A := \boldsymbol{s}_A + \frac{1}{c^2} \left[\boldsymbol{s}_A \hat{\mathscr{U}}_{\neg A}\left(t,\boldsymbol{x}_A\right) +\frac{1}{2} \left(\boldsymbol{s}_A \cdot \boldsymbol{v}_A\right)\boldsymbol{v}_A \right],
\label{eq:refined-spin}
\end{align}
and write Eq. \eqref{eq:spin-precession-1} as
\begin{align}
\frac{\dd \bar{\boldsymbol{s}}_A}{\dd t}    = \boldsymbol{\Omega}_A \times  \bar{\boldsymbol{s}}_A+ \OO\left(c^{-4}\right), 
\label{eq:spin-precession-2}    
\end{align}
where $\boldsymbol{\Omega}_A^{\rm SS}$ can be easily written in terms of $\bar{\boldsymbol{s}}_A$ bearing in mind that $ \bar{\boldsymbol{s}}_A = \boldsymbol{s}_A  + \OO\left(c^{-2}\right)$. The new spin vector \eqref{eq:refined-spin} slightly differs from the analogous redefined angular momentum adopted in GR (see Sect. 9.5.1 in Ref. \cite{Poisson-Will2014}). In fact, in EC framework no terms related to the translational kinetic energy occur, while the corrections coming from the internal structure of the body  are included in the PN series \eqref{eq:PN-vector-s-A}\footnote{In our model, these internal corrections do not depend on the  velocity $\boldsymbol{w}$ of each fluid element of the body relative to $\boldsymbol{v}_A$, since  $\boldsymbol{w}$ vanishes due to the hypothesis of stationary equilibrium.}. 

Although our starting point is represented by the EC-framed Eq. \eqref{eq:rotational_fluid_equation}, it is clear that Eq. \eqref{eq:spin-precession-2} reproduces the corresponding GR equations pertaining to the evolution of  the macroscopic angular momentum if $\bar{\boldsymbol{s}}_A$ is multiplied by a factor 2.  This represents an  important consistency check of the EC model, since the same conclusion holds also for the translational motion \cite{Paper3}.  We stress that, in order to obtain this result, the role of the Frenkel condition \eqref{eq:Frenkel_condition} is crucial. Furthermore,  it is worth pointing out that, likewise the case of the translational dynamics, the internal components of the bodies do not give contribution to the rotational motion  \eqref{eq:spin-precession-2}. In fact, the term $2 \varepsilon_{ijp} \mathcal{Z}_A^{[ji]}$, originating from Eq. \eqref{eq:PPL-spin-equation-1}, is vanishing at 1PN order upon performing the involved integrations (a detailed calculation is given in  Appendix \ref{Sec:Appendix-integral-Z}). This underlines, once again, the validity of the effacing principle at 1PN order in EC theory endowed with the physical condition $S^{\alpha\mu}{}_\mu=0$.

\subsection{Physical interpretation of the inner-structure-dependent quantities}
\label{Sec:inner-structure-dependent quantities}
In this section, we discuss the physical interpretation of the following  inner-structure-dependent quantities:
\begin{subequations}
\begin{align}
\mathcal{H}^{ki}_A&:= 3G\int_A \dd^3 \boldsymbol{y} \, \dd^3 \boldsymbol{y}^\prime \, \rho^\star s^\prime_{kj} \frac{(y-y^\prime)^{\langle i}(y-y^\prime)^{j \rangle}}{\vert \boldsymbol{y}-\boldsymbol{y}^\prime \vert^5},\label{eq:tensor-mathcal-H-A-ki}\\
\mathcal{Q}^{ij}_A&:= 6G\int_A \dd^3 \boldsymbol{y} \, \dd^3 \boldsymbol{y}^\prime \, s_{ik} s^\prime_{lj} \frac{(y-y^\prime)^{\langle k}(y-y^\prime)^{l \rangle}}{\vert \boldsymbol{y}-\boldsymbol{y}^\prime \vert^5},\label{eq:tensor-mathcal-Q-A-ij}\\
\mathcal{Z}^{ij}_A&:= 6G\int_A \dd^3 \boldsymbol{y} \, \dd^3 \boldsymbol{y}^\prime \, s_{ik} s^\prime_{lk} \frac{(y-y^\prime)^{\langle l}(y-y^\prime)^{j \rangle}}{\vert \boldsymbol{y}-\boldsymbol{y}^\prime \vert^5},\label{eq:tensor-mathcal-Z-A-ij}
\end{align}
\end{subequations}
where $y^i := x^i -x^i_A\left(t\right)$. Note that  Eq. \eqref{eq:tensor-mathcal-H-A-ki} occurs in the computations of the translational motion (see Ref. \citep{Paper3}, for details), while Eqs. \eqref{eq:tensor-mathcal-Q-A-ij} and \eqref{eq:tensor-mathcal-Z-A-ij} appear in the rotational dynamics \eqref{eq:PPL-spin-equation-1}. 

If we perform a dimensional analysis of the above terms, we obtain  
\begin{subequations}
\begin{align}
\left[\mathcal{H}^{ki}_A \right]&=\frac{{\rm mass}\cdot({\rm length})^2}{({\rm time})^3},\label{eq:dim-H-A-ki}\\
\left[\mathcal{Q}^{ij}_A \right]&=\left[\mathcal{Z}^{ij}_A \right]=\frac{{\rm mass}\cdot({\rm length})^4}{({\rm time})^4}.\label{eq:dim-QeZ-A-ij}
\end{align}
\end{subequations}
Therefore, the following interpretations are in order: (1) $\mathcal{H}^{ki}_A$ may be seen as the second-time variation of the spin inside the body; (2) $\mathcal{Q}^{ij}_A$ and $\mathcal{Z}^{ij}_A$ may resemble the second-time variation of the spin quadrupole-like effects inside the body. The subtle differences between $\mathcal{Q}^{ij}_A$ and $\mathcal{Z}^{ij}_A$ rely on the saturation of the indices inside the integrals. We see that
$\mathcal{Q}^{ij}_A$ and $\mathcal{Z}^{ij}_A$ represent the spin-body-body and spin-spin-body interactions, respectively.

\section{Lagrangian and first integrals for a binary system}
\label{sec:Lagrangian-EC-theory}
In the previous section, we have investigated the 1PN-accurate dynamics of an $N$-body system.  In this section, we restrict our attention to binary systems and  provide a Lagrangian formulation for their dynamics.

In GR, the (conservative) harmonic-coordinate equations of motion of a two-body system can be derived from a generalized Lagrangian, which besides the (relative) position and velocity vectors, depends also on the (relative) acceleration. This effect occurs  at 2PN level if the objects have no angular momentum, or already at 1PN order in the case they  have a \qm{classic spin} \citep{Blanchet2014}. This result is formally justified by the theorem of  Martin and Sanz, which is valid as long as the adopted gauge conditions  are  Lorentz invariant \cite{Martin1979}. The acceleration dependence in the Lagrangian can be obtained by means of the previous PN-expanded equations of motion and can  always be recast in a linear form via the addition of the so-called \emph{multi-zero terms}  \cite{Blanchet2001,Blanchet2014}. In general, the occurrence of the acceleration in the  Lagrangian is obtained via a guess-work procedure \cite{Damour1983book}. A way out of this issue consists in the use of \emph{contact transformations} and Arnowitt, Deser, and Misner (ADM) coordinates, which permit to recover an ordinary Lagrangian \citep{Blanchet2014}. Indeed,  this strategy does not violate  Martin and Sanz theorem, because the ADM coordinate conditions break the Lorentz invariance \cite{Damour1985b}. In the current literature, which is devoted to the description of binary dynamics at high PN orders, the most common approach relies on the Hamiltonian formalism in ADM coordinates, which avoid the occurrence of accelerations \cite{Schafer2014,Schafer2018}. 

As we will see in this section, the same situation as in the GR framework occurs also in EC theory, as (the SO part of) the Lagrangian involves acceleration terms. After having derived the equations of motion of the two-body system in Sect. \ref{sec:2B-eqs-motion}, the Lagrangian and the ensuing first integrals will be computed in Secs. \ref{Sec:2B-Lagrangian} and \ref{Sec:first-integrals}, respectively. 

\subsection{Two-body equations of motion}\label{sec:2B-eqs-motion}
By eliminating the center of mass of the system \cite{Paper2,Paper3}, the two-body problem admits an  effective one-body description whose main variables are represented by the following relative vectors:
\begin{align}\nonumber
    &\boldsymbol{r}:=\boldsymbol{x}_1-\boldsymbol{x}_2, \qquad\qquad\quad \boldsymbol{n} := \boldsymbol{r}/r, \\
    & \boldsymbol{v}:= \dfrac{\dd}{\dd t}\boldsymbol{r} = \boldsymbol{v}_1 - \boldsymbol{v}_2, \qquad \, \boldsymbol{a}:= \dfrac{\dd}{\dd t}\boldsymbol{v} = \boldsymbol{a}_1 - \boldsymbol{a}_2.
\end{align}
In our forthcoming analysis, it is  also useful to introduce the spin variables 
\begin{align}
    \boldsymbol{s}:= \boldsymbol{s}_1 + \boldsymbol{s}_2, \qquad \boldsymbol{\sigma}:= \dfrac{M_2}{M_1}\boldsymbol{s}_1 + \dfrac{M_1}{M_2}\boldsymbol{s}_2,
\end{align}
and the total mass $M$, the reduced mass $\mu$, and the symmetric mass ratio $\nu$ of the system
\begin{align}
    M := M_1 + M_2, \qquad \mu := \dfrac{M_1 M_2}{M}, \qquad \nu := \dfrac{\mu}{M}.
\end{align}

The translational dynamics of the two-body system can be described at 1PN level by means of the relative acceleration (cf. Eq. \eqref{eq:EC-body-A-equation-of-motion-2})
\begin{align}
\label{eq:acceleration-complete}
\boldsymbol{a}=\underbrace{\boldsymbol{a}_{\rm N}+\boldsymbol{a}_{\rm 1PN}}_{\boldsymbol{a}_{\rm EIH}}+\underbrace{\boldsymbol{a}_{\rm SO}+\boldsymbol{a}_{\rm SS}}_{\boldsymbol{a}_{\rm EC}} + \OO\left(c^{-4}\right),
\end{align}
where the GR contribution is \cite{Damour1985,Kidder1993}
\begin{subequations}
\begin{align}
\boldsymbol{a}_{\rm N}&=-\frac{G M}{r^2}\boldsymbol{n},\\
\boldsymbol{a}_{\rm 1PN}&=\frac{G M}{c^2 r^2} \Biggr{\{}\boldsymbol{n}\left[\frac{2G M}{r}( \nu +2)-(3 \nu +1) v^2\right.\notag\\
&\left.+\frac{3}{2} \nu  (\boldsymbol{n}\cdot\boldsymbol{v})^2\right]+2(2- \nu ) (\boldsymbol{n}\cdot\boldsymbol{v})\boldsymbol{v}\Biggr{\}},
\end{align}
\end{subequations}
 while the EC correction reads as \cite{Paper3}
\begin{subequations}
\label{eq:acceleration-SO-SS}
\begin{align}
\boldsymbol{a}_{\rm SO}&=2\Biggr{\{}\frac{G}{c^2 r^3}\Biggr{[}3 (\boldsymbol{n}\cdot\boldsymbol{v})(\boldsymbol{n}\times(\boldsymbol{\sigma} +2 \boldsymbol{s}))\notag\\
&+6\boldsymbol{n} (\boldsymbol{n}\times \boldsymbol{v})\cdot(\boldsymbol{\sigma}+ \boldsymbol{s})-\boldsymbol{v}\times (3\boldsymbol{\sigma} +4 \boldsymbol{s})\Biggr{]}\Biggr{\}},\\
\boldsymbol{a}_{\rm SS}&=4\Biggr{\{}\frac{3 G}{c^2 \mu  r^4}\Biggr{[} 5 \boldsymbol{n} (\boldsymbol{n}\cdot\boldsymbol{s}_1)( \boldsymbol{n}\cdot\boldsymbol{s}_2)-\boldsymbol{s}_1(\boldsymbol{n}\cdot\boldsymbol{s}_2)\notag\\
&-\boldsymbol{s}_2 (\boldsymbol{n}\cdot\boldsymbol{s}_1)-\boldsymbol{n} (\boldsymbol{s}_1\cdot\boldsymbol{s}_2)\Biggr{]}\Biggr{\}}.
\end{align}
\end{subequations}
Note that $\boldsymbol{a}_{\rm SO}$ and $\boldsymbol{a}_{\rm SS}$ are proportional to the GR accelerations by the factors highlighted outside  the curly brackets in  Eq. \eqref{eq:acceleration-SO-SS}. This result ties in with our comment below Eq. \eqref{eq:spin-precession-2}.

It follows from the outcome of Sect. \ref{Sec:rotational-motion}, that for binary systems the rotational motion 
\begin{align}
    \frac{\dd \bar{\boldsymbol{s}}_A}{\dd t} = \left( \boldsymbol{\Omega}_A^{\rm SO} +  \boldsymbol{\Omega}_A^{\rm SS} \right) \times \boldsymbol{s}_A + \OO\left(c^{-4}\right), \qquad (A=1,2),
\end{align}
is governed by  the precessional angular velocities 
\begin{subequations}
\label{eq:precessional-velocities-EC}
\begin{align}
\boldsymbol{\Omega}_1^{\rm SO} &=\Biggr{\{} \frac{2G \mu}{c^2r^2}\left(1+\frac{3M_2}{4M_1}\right) (\boldsymbol{n}\times \boldsymbol{v})\Biggr{\}},\label{eq:Omega-SO_1}\\ 
\boldsymbol{\Omega}_2^{\rm SO} &=\Biggr{\{} \frac{2G \mu}{c^2r^2}\left(1+\frac{3M_1}{4M_2}\right) (\boldsymbol{n}\times \boldsymbol{v})\Biggr{\}},\label{eq:Omega-SO_2}\\ 
\boldsymbol{\Omega}_1^{\rm SS} &=2\Biggr{\{} \frac{G}{c^2r^3}\Biggr{[} 3 \left(\boldsymbol{n} \cdot \boldsymbol{s}_2\right)\boldsymbol{n} -\boldsymbol{s}_2 \Biggr{]}\Biggr{\}},
\label{eq:Omega-SS_1} \\
\boldsymbol{\Omega}_2^{\rm SS} &=2\Biggr{\{} \frac{G}{c^2r^3}\Biggr{[} 3 \left(\boldsymbol{n} \cdot \boldsymbol{s}_1\right)\boldsymbol{n} -\boldsymbol{s}_1 \Biggr{]}\Biggr{\}},
\label{eq:Omega-SS_2}    
\end{align}
\end{subequations}
where the curly brackets make it clear that    $\boldsymbol{\Omega}_A^{\rm SO}$ assumes the same form as in GR, whereas $\boldsymbol{\Omega}_A^{\rm SS}$ is twice its GR counterpart.   

\subsection{Lagrangian  formulation}
\label{Sec:2B-Lagrangian}
As pointed out before, the translational dynamics  can be formulated in terms of an acceleration-dependent Lagrangian. In fact,  bearing in mind the GR results \cite{Damour1985,Kidder1993}, we find that the 1PN-accurate Lagrangian function of the binary system is
\begin{equation} \label{eq:Lagrangian_EC_theory}
\mathcal{L}\left(\boldsymbol{r},\boldsymbol{v},\boldsymbol{a}\right)=\underbrace{\mathcal{L}_{\rm N}+\mathcal{L}_{\rm 1PN}}_{\mathcal{L}_{\rm GR}}+\underbrace{\mathcal{L}_{\rm SO}+\mathcal{L}_{\rm SS}}_{\mathcal{L}_{\rm EC}} + \OO\left(c^{-4}\right),    
\end{equation}
where the GR piece is
\begin{subequations}
\begin{align}
\mathcal{L}_{\rm N}&=\mu\left(\frac{ v^2}{2}+\frac{G M }{r}\right),\\
\mathcal{L}_{\rm 1PN}&=\frac{\mu}{c^2}\Biggr{\{}\frac{G M}{2r} \left[-\frac{G M}{r}+\nu  (\boldsymbol{n}\cdot\boldsymbol{v})^2+(\nu +3) v^2\right]\notag\\
&\quad\qquad+\frac{1}{8} (1-3 \nu ) v^4\Biggr{\}},
\end{align}
\end{subequations}
while the EC term is given by
\begin{equation}\label{eq:LagrangianEC-SO-SS}
\mathcal{L}_{\rm EC}=-2\sum_{A=1}^2 \left(\boldsymbol{\Omega}^{\rm SO}_A+\frac{1}{2}\boldsymbol{\Omega}^{\rm SS}_A\right)\cdot\boldsymbol{s}_A,
\end{equation}
and hence reads as (cf. Eq. \eqref{eq:precessional-velocities-EC})
\begin{subequations}
\label{eq:L-SO&L-SS}
\begin{align}
\mathcal{L}_{\rm SO}&=2\Biggr{\{} \frac{\mu}{c^2}\left[\frac{ \boldsymbol{v}\cdot(\boldsymbol{a}\times \boldsymbol{\sigma}) }{2 M}+\frac{2G}{ r^3} \boldsymbol{v}\cdot(\boldsymbol{r}\times (\boldsymbol{\sigma}+ \boldsymbol{s}))\right]\Biggr{\}},
\label{eq:L-SO}\\
\mathcal{L}_{\rm SS}&=4\Biggr{\{}\frac{G}{c^2 r^3} \Biggr{[}\boldsymbol{s}_1\cdot\boldsymbol{s}_2-3(\boldsymbol{n}\cdot\boldsymbol{s}_1)( \boldsymbol{n}\cdot\boldsymbol{s}_2)\Biggr{]}\Biggr{\}}.
\label{eq:L-SS}
\end{align}
\end{subequations}
We stress that $\mathcal{L}_{\rm SO}$ and $\mathcal{L}_{\rm SS}$ reproduce their GR analogues if the spin vector $\boldsymbol{s}_A$ is divided by a factor 2. 

The translational equations of motion \eqref{eq:acceleration-complete} stem from the Euler-Lagrange equations 
\begin{subequations}
\begin{align}
\boldsymbol{0}&=\frac{\partial \mathcal{L}}{\partial \boldsymbol{r}}-\frac{\dd \boldsymbol{p}}{\dd t},\\
\boldsymbol{p}&=\frac{\partial \mathcal{L}}{\partial \boldsymbol{v}}-\frac{\dd \boldsymbol{j}}{\dd t},\\
\boldsymbol{j}&=\frac{\partial \mathcal{L}}{\partial \boldsymbol{a}},
\end{align}
\end{subequations}
$\boldsymbol{p}$  being  the generalized canonical momentum.

The rotational dynamics can be easily dealt with if we resort to the Hamiltonian formalism. Within this pattern, the SO and SS couplings are described by the Hamiltonian function
\begin{align} \label{eq:Hamiltonian-EC-theory}
\mathcal{H}(\boldsymbol{r},\boldsymbol{\mathcal{P}},\boldsymbol{s}_1,\boldsymbol{s}_2)=\mathcal{H}_{\rm SO} + \mathcal{H}_{\rm SS}    + \OO\left(c^{-4}\right),
\end{align}
with 
\begin{subequations}
\begin{align} 
\mathcal{H}_{\rm SO} &=2\Biggr{\{}\frac{G \mu}{c^2 r^3} \left[\left(\boldsymbol{r}\times \frac{\boldsymbol{\mathcal{P}}}{\mu}\right)\cdot\left(\frac{3 }{2}\boldsymbol{\sigma}+2\boldsymbol{s} \right)\right] \Biggr{\}},\\
\mathcal{H}_{\rm SS} &= 4\Biggr{\{}\frac{G}{c^2 r^3} \Biggr{[}3(\boldsymbol{n}\cdot\boldsymbol{s}_1)( \boldsymbol{n}\cdot\boldsymbol{s}_2)-\boldsymbol{s}_1\cdot\boldsymbol{s}_2\Biggr{]}\Biggr{\}}, 
\end{align}
\end{subequations}
and $\boldsymbol{\mathcal{P}}=\mu \boldsymbol{v}$  the (relative) kinematic  momentum (see e.g. Refs. \cite{Barker1975,Damour2001}, for further details). By employing the  expression of  $\boldsymbol{\mathcal{P}}$ and the PN formula $\boldsymbol{a}=\boldsymbol{a}_N + \OO\left(c^{-2}\right)$,  one can write $\mathcal{H} = - \left(\mathcal{L}_{\rm SO} + \mathcal{L}_{\rm SS}\right)+ \OO\left(c^{-4}\right)$  (see Eqs. \eqref{eq:acceleration-complete} and \eqref{eq:L-SO}). We note that  the replacement of  the acceleration by its Newtonian value in the 1PN generalized Lagrangian is a correct procedure only when we cope with the spin motion. 

The Hamiltonian approach permits to characterize the spin precession via the Poisson brackets $\{\cdot,\cdot\}$  as
\begin{align}
\frac{\dd \bar{\boldsymbol{s}}_A}{\dd t}&=\left\{\boldsymbol{s}_A,\mathcal{H}\right\}
=\frac{1}{2}\left\{\frac{\partial \mathcal{H}}{\partial \boldsymbol{s}_A}\times \boldsymbol{s}_A \right\}, \qquad (A=1,2),
\end{align}
upon  exploiting the basic relations 
\begin{equation}
\left\{s_A^i,s_A^j\right\}=\varepsilon_{ijk}s_A^k,\qquad (A=1,2),
\end{equation}
and the fact that the orbital variables $\boldsymbol{r}$ and $\boldsymbol{\mathcal{P}}$ have vanishing Poisson brackets with  the spin variables.

The above equations can be also written in terms of the Lagrangian \eqref{eq:Lagrangian_EC_theory} as follows 
\begin{align}
\frac{\dd \bar{\boldsymbol{s}}_A}{\dd t}&=\frac{1}{2}\left\{\boldsymbol{s}_A \times \frac{\partial \mathcal{L}}{\partial \boldsymbol{s}_A}\right\},\qquad (A=1,2).
\end{align}
Therefore, the 1PN rotational dynamics pertaining to the total refined spin tensor (see Eq. \eqref{eq:refined-spin})
\begin{align}
    \bar{\boldsymbol{s}} = \bar{\boldsymbol{s}}_1 + \bar{\boldsymbol{s}}_2,
\end{align}
is represented by
\begin{align}
\frac{\dd \bar{\boldsymbol{s}}}{\dd t}&=\frac{G}{c^2 r^3}\Biggr{\{}  \boldsymbol{L}_{\rm N}\times \left(\frac{3 \boldsymbol{\sigma}}{2}+2 \boldsymbol{s} \right)+6\Biggr{[}(\boldsymbol{n}\cdot\boldsymbol{s}_1)(\boldsymbol{n}\times \boldsymbol{s}_2)\notag\\
&+(\boldsymbol{n}\cdot\boldsymbol{s}_2)(\boldsymbol{n}\times \boldsymbol{s}_1)\Biggr{]}\Biggr{\}} + \OO\left(c^{-4}\right),
\label{eq:spin-precession-total-spin}
\end{align}
the Newtonian angular momentum being
\begin{align}
\boldsymbol{L}_{\rm N}&=\mu ( \boldsymbol{r}\times \boldsymbol{v}).  
\end{align}

In the above analysis,  the study of spin motion relies upon the Hamiltonian picture, while the Lagrangian pattern has been deduced only in a second moment. Despite that, it is possible to 
investigate the spin precession by exploiting exclusively the Lagrangian approach. Indeed, in Refs. \cite{Barker1970,Barker1975}, it has been shown that the GR rotational dynamics can be derived from a Lagrangian formalism  provided that the motion is described in terms of the Euler angles and the rotational kinetic energy of the system is added to the Lagrangian function. Although the GR framework concerns the evolution of the  angular momentum and not of the quantum spin, it is possible to construct a similar scheme also in EC theory. The main steps are the following. First of all, since our investigation is restricted to 1PN results, it makes sense to resort to nonrelativistic quantum mechanics. Moreover, to fix the ideas, let us consider  spin-1/2 particles. It is known that their analysis rests on the two-dimensional spinorial representation of  $SU(2)$, which is the double cover of the rotation group $SO(3)$ \cite{Esposito2002,Maggiore2005}. Since the elements of an orthogonal transformation can be expressed in terms of the three Euler angles, these can be exploited also in EC model to describe the spin precession via the Lagrangian formulation. Finally, the spin kinetic energy of each body, which should be added to the Lagrangian function, can be constructed starting from the  spin kinetic energy density of the Weyssenhoff fluid 
\begin{equation}\label{eq:spin-energy}
K_{\rm spin} = \frac{1}{2}s^{\mu \nu}\omega_{\mu \nu},
\end{equation}
$\omega_{\mu\nu}$ being the fluid microscopic angular velocity (see the discussion regarding the first thermodynamic law in Ref. \cite{Paper2}, for further details). 

\subsection{First integrals} 
\label{Sec:first-integrals}
Having obtained the Lagrangian formulation of the 1PN dynamics of the binary system, the  first integrals can be easily computed. Indeed, the total  energy  reads as 
\begin{equation} \label{eq:ENERGY}
E=\boldsymbol{p}\cdot\boldsymbol{v}+\boldsymbol{a}\cdot\boldsymbol{j}-\mathcal{L},
\end{equation}
and its  full expression is
\begin{equation}
E=\underbrace{E_{\rm N}+E_{\rm 1PN}}_{E_{\rm GR}}+\underbrace{E_{\rm SO}+E_{\rm SS}}_{E_{\rm EC}}+ \OO\left(c^{-4}\right),    
\end{equation}
where
\begin{subequations}
\begin{align}
E_{\rm N}&=\mu\left(\frac{v^2}{2}-\frac{G M}{r}\right),\\
E_{\rm 1PN}&=\frac{\mu}{c^2}\Biggr{\{}\frac{G M}{2 r} \left[\frac{G M}{r}+\nu  (\boldsymbol{n}\cdot\boldsymbol{v})^2+(\nu +3) v^2\right]\notag\\
&+\frac{3}{8}(1-3 \nu ) v^4\Biggr{\}},\\
E_{\rm SO}&=2\Biggr{\{}\frac{G\mu}{c^2 r^2} (\boldsymbol{n}\times\boldsymbol{v})\cdot \boldsymbol{\sigma} \Biggr{\}},\\
E_{\rm SS}&=4\Biggr{\{}\frac{G}{c^2 r^3}\Biggr{[}3(\boldsymbol{n}\cdot\boldsymbol{s}_1)( \boldsymbol{n}\cdot\boldsymbol{s}_2)-\boldsymbol{s}_1\cdot\boldsymbol{s}_2\Biggr{]}\Biggr{\}}.
\end{align}
\end{subequations}
Moreover, the total angular momentum of the system is
\begin{equation} \label{eq:TOTAL_ANGULAR_MOMENTUM}
\boldsymbol{J}=(\boldsymbol{r}\times\boldsymbol{p})+(\boldsymbol{v}\times\boldsymbol{j})+\bar{\boldsymbol{s}},  
\end{equation}
and it  can be explicitly written as
\begin{equation}
\boldsymbol{J}=\underbrace{\boldsymbol{L}_{\rm N}+\boldsymbol{L}_{\rm 1PN}}_{\boldsymbol{L}_{\rm GR}}+\underbrace{\boldsymbol{L}_{SO}}_{\boldsymbol{L}_{\rm EC}}+\bar{\boldsymbol{s}}+ \OO\left(c^{-4}\right),    
\end{equation}
where
\begin{subequations}
\begin{align}
\boldsymbol{L}_{\rm 1PN}&=\frac{\boldsymbol{L}_{\rm N}}{c^2} \left[\frac{G M}{r}(\nu +3)+(1-3 \nu )\frac{ v^2}{2}\right]
,\\
\boldsymbol{L}_{SO}&=2\Biggr{\{}\frac{ \mu}{c^2 M}  \Biggr{[}\frac{G M}{r} \boldsymbol{n}\times (\boldsymbol{n}\times (\boldsymbol{\sigma} +2 \boldsymbol{s} ))\notag\\
&\hspace{1.8cm}-\frac{1}{2} \boldsymbol{v}\times (\boldsymbol{v}\times \boldsymbol{\sigma} )\Biggr{]}\Biggr{\}}.
\end{align}
\end{subequations}
Note that, in the above equations,  $E_{\rm SO}$, $E_{\rm SS}$, and $\boldsymbol{L}_{SO}$  are proportional to their corresponding GR quantities \cite{Kidder1993}, as it should be expected.

By exploiting Eqs. \eqref{eq:acceleration-complete} and \eqref{eq:spin-precession-total-spin}, it is easy to show that the motion keeps $E$ and $\boldsymbol{J}$ constant, i.e.,  $\dd E / \dd t=0$ and $\dd \boldsymbol{J} / \dd t=0$.

\section{Conclusions} 
\label{sec:end}
In this paper, we have worked out the rotational motion at 1PN order of an $N$-body system in EC theory and, for the special case of binary systems, we have provided the Lagrangian formulation 
and the first integrals governing the dynamics.

In  Sect. \ref{sec:N-body-rotational}, the 1PN spin precession equations have been obtained by applying the point-particle procedure to the rotational motion of the Weyssenhoff fluid, which is the model we have adopted to describe the quantum spin effects occurring inside the bodies. Unlike the translational dynamics, where the contributions coming from  inner-structure-dependent integrals cancel algebraically, the rotational motion \eqref{eq:PPL-spin-equation-1} is characterized by the structure term  $2\varepsilon_{ijp} \mathcal{Z}_A^{[ji]}$ which does not vanish in a trivial way. This fact should be expected  on general grounds, as the integral $\mathcal{Z}_A^{ji}$ embodies the zero-range spin interaction which represents the new original feature brought in  by EC theory (see Sect.  \ref{Sec:inner-structure-dependent quantities}). However, after a detailed investigation (see Appendix \ref{Sec:Appendix-integral-Z}), we have proved that   $2\varepsilon_{ijp} \mathcal{Z}_A^{[ji]}$ amounts to zero, provided that the reflection symmetry hypothesis is taken into account. This remarkable result entails a twofold implication: on the one hand, the equations of motion have the same functional form as in GR (up to a  multiplicative factor in the spin) and, on the other, we can claim that \emph{effacing principle} is valid at 1PN order in EC theory (at least for matter models enforcing the   condition $S^{\alpha\mu}{}_\mu=0$, see paragraph below Eq. \eqref{eq:Frenkel_condition}). Driven by these arguments and  the resemblance to GR, it has been easy in Sect. \ref{sec:Lagrangian-EC-theory} to derive the Lagrangian function and the first integrals ruling the 1PN dynamics of a binary system. The characterization of the spin precession within the Lagrangian picture requires the introduction of the Euler angles, which can be easily defined also in EC theory if we exploit the two-to-one homomorphism between  $SU(2)$ and $SO(3)$. 

We have already discussed in Ref. \cite{Paper3} that the deviations from the GR bulk dynamics turn out to be very tiny. These have been evaluated by supposing that all the elementary spins inside the bodies are aligned along a preferred direction and hence they represent the largest corrections introduced by EC model. Indeed, we recall that for unpolarized matter the terms which are linear in the spin or involve its gradient vanish upon performing a spacetime averaging procedure, while  quadratic-in-spin factors, which are distinct ingredients of EC theory, give a nonzero contribution \cite{Hehl1976_fundations,Gasperini1986,Hashemi2015,Medina2018}. However, there exist in nature configurations where the alignment of the spins naturally occurs as a consequence of the presence of some external polarizing field. In fact, in the case of neutron stars, the strong magnetic fields, together with the spin-torsion forces and the strong-gravity interaction yield this alignment (see Refs. \cite{deSabbata1980,Chamel2008}, for more details). On the other hand, for black holes we assume that the spins are aligned, since we have no insight into their inner structure. This represents a first approach, which permits to probe possible spin effects in black-hole physics by means of e.g., GW phenomena. Indeed, more refined models could potentially shed light on black hole interior and quantum-gravity issues.

This paper, along with the previous works \cite{Paper1, Paper2, Paper3}, constitutes a comprehensive examination of the GW generation problem and the $N$-body dynamics at the 1PN order in EC theory via the Weyssenhoff fluid. Our research program opens up several interesting perspectives for future studies, such as: $(i)$  determining the analytical solution of the translational motion \eqref{eq:acceleration-complete}, similarly to what has been done  in GR for bodies with no angular momentum \cite{Damour1985};  $(ii)$  deriving  the equations of motion of a two-body system at 2PN level to check whether some differences with respect to GR would emerge; $(iii)$  employing  a different model from the Weyssenhoff semiclassical one to explore the spin effects and possible deviations from GR at  various PN orders; $(iv)$ considering some  applications of our findings to astrophysical settings as well as their generalization to cosmology, where EC theory can lead to interesting implications, as the recent literature shows
(see e.g., Refs. \cite{Medina2018,Benisty2021,Elizalde2022,Pereira2022}). 

\section*{Acknowledgements}
E.B. and V.D.F are grateful to  Gruppo Nazionale di Fisica Matematica of Istituto Nazionale di Alta Matematica for support. V.D.F. and D.U. acknowledge the support of INFN {\it sez. di Napoli}, {\it iniziative specifiche} TEONGRAV and QGSKY. E.B. acknowledges the support of  the Austrian Science Fund (FWF) grant P32086. 

\appendix
\section{$N$-body problem in GR theory} 
\label{sec:GR-Appendix}
In this appendix, we briefly recall the $N$-body problem at 1PN level in GR. 

The Einstein field equations, when solved iteratively via the PN method and the \emph{harmonic gauge condition} are transformed into a set of Poisson equations, which, at 1PN order, can be solved in terms of the following \emph{instantaneous potentials}  \cite{Maggiore:GWs_Vol1,Blanchet2014}: 
\begin{subequations}
\label{eq:inst_potentials}
\begin{align}
\phi(t,\boldsymbol{x})&:=-\frac{G}{c^4}\int {\rm d}^3\boldsymbol{x}^\prime\,\frac{{}^{(0)}T^{00}(t,\boldsymbol{x}^\prime)}{|\boldsymbol{x}-\boldsymbol{x}^\prime|},
\label{eq:phi-potential}
\\
\zeta_i(t,\boldsymbol{x})&:=-\frac{4G}{c^4}\int{\rm d}^3\boldsymbol{x}^\prime\, \frac{{}^{(1)}T^{0i}(t,\boldsymbol{x}^\prime)}{|\boldsymbol{x}-\boldsymbol{x}^\prime|},
\label{eq:zeta-potential}
\\
\xi(t,\boldsymbol{x})&:=-\frac{1}{2c^4}\partial^2_{t}\chi(t,\boldsymbol{x})
\nn \\
&-\frac{G}{c^4}\int {\rm d}^3\boldsymbol{x}^\prime\,\frac{{}^{(2)}T^{00}(t,\boldsymbol{x}^\prime)+{}^{(2)}T^{ii}(t,\boldsymbol{x}^\prime)}{|\boldsymbol{x}-\boldsymbol{x}^\prime|},
\label{eq:psi_potential}
\end{align}
\end{subequations}
where the stress-energy tensor is such that ${}^{(n)}T^{\mu \nu} = \OO \left(c^{2-n}\right)$. The function  $\chi(t,\boldsymbol{x})$ occurring in Eq. \eqref{eq:psi_potential} is refereed to as  \emph{superpotential}. It fulfils a crucial role in the evaluation  of the integral expressions involving the time  derivatives of $\phi$.

In the case of a system consisting of $N$ point-like particles having masses $m_A$ and moving along  trajectories described by the relations $\boldsymbol{x}=\boldsymbol{x}_A(t)$ and with velocity $\boldsymbol{v}_A$, the instantaneous potentials \eqref{eq:inst_potentials} can be written as  ($\boldsymbol{d}_A :=  \boldsymbol{x}-\boldsymbol{x}_A $, $\boldsymbol{n}_A := \boldsymbol{d}_A/d_A $)
\cite{Maggiore:GWs_Vol1,Blanchet2014,Poisson-Will2014}
\begin{subequations}
\label{eq:inst-potential-N-body}
\begin{align}
\phi&=-\dfrac{G}{c^2}\sum_{A}\frac{m_A}{d_A},
\\
\zeta_i&=-\dfrac{4G}{c^3}\sum_{A}\frac{m_A v_A^i}{d_A},
\\
\xi&= -\dfrac{2G}{c^4}\sum_{A} \dfrac{m_A \boldsymbol{v}^2_A}{d_A}+\dfrac{G}{2c^4}\sum_{A} \dfrac{m_A \left(\boldsymbol{v}_A\cdot \boldsymbol{n}_A\right)^2}{d_A} 
\nn \\
&+\dfrac{G}{2c^4} \sum_A m_A \boldsymbol{n}_A \cdot \boldsymbol{a}_A + \dfrac{G^2}{c^4} \sum_A \sum_{B \neq A} \dfrac{m_A m_B}{d_A r_{AB}},
\label{eq:potential-psi-N-body}
\end{align}
\end{subequations}
where in deriving Eq. \eqref{eq:potential-psi-N-body}  we have exploited the regularization prescription (which is a special case of Hadamard regularization \cite{Blanchet2014}) 
\begin{align}\label{eq:regular-prescription}
    \dfrac{\delta^{(3)}\left(\boldsymbol{d}_A\right)}{d_A} \equiv 0,
\end{align}
to work out the otherwise ill-defined integral (see Eq. \eqref{eq:phi-potential})
\begin{align}
    \int   \dfrac{\dd^3 \boldsymbol{x}^\prime}{\vert \boldsymbol{x}-\boldsymbol{x}^\prime \vert}\,\phi(t,\boldsymbol{x}^\prime) \delta^{(3)}\left(\boldsymbol{x}^\prime-\boldsymbol{x}_A(t)\right);
\end{align}
we refer the reader to Sect. 9.6 in Ref. \cite{Poisson-Will2014}, for further details.  

The Lagrangian function pertaining to the geodesic motion of the $N$-body system  reads as
\begin{align}\label{eq:GR-Lagrangian-N-Body}
L^{N{\rm B}} :=  L^{N{\rm B}}_{0}+\frac{1}{c^2}L^{N{\rm B}}_{2} + \OO(c^{-4}), 
\end{align}
with 
\begin{subequations}
\label{eq:N-body-Lagrangian-GR-1PN}
\begin{align} \label{eq:GR-Newtonian-Lagrangian-N-Body}
L^{N{\rm B}}_{0} &:= \dfrac{1}{2}\sum_A m_A v_A^2 + \dfrac{G}{2}\sum_A \sum_{B \neq A} \dfrac{m_A m_B}{r_{AB}},
\\ 
L^{N{\rm B}}_{2}&:= \dfrac{1}{8}\sum_A m_A v_A^4 -\dfrac{G}{4} \sum_A \sum_{B \neq A} \dfrac{m_A m_B}{r_{AB}} \Bigl[7 \boldsymbol{v}_A \cdot \boldsymbol{v}_B 
\nn \\
&+ \left(\boldsymbol{v}_A \cdot \boldsymbol{n}_{AB}\right)\left(\boldsymbol{v}_B \cdot \boldsymbol{n}_{AB}\right)\Bigr]+\dfrac{3G}{2} \sum_A \sum_{B \neq A} \dfrac{m_A m_B}{r_{AB}}v_A^2
\nn \\
&-\dfrac{G^2}{2} \sum_A \sum_{B \neq A}\sum_{C \neq A} \dfrac{m_A m_B m_C}{r_{AB}r_{AC}},
\label{eq:GR-1PN-Lagrangian-N-Body}
\end{align}
\end{subequations}
where in the derivation of Eq.  \eqref{eq:GR-1PN-Lagrangian-N-Body} we have discarded a total time derivative. It is important to note that in  Eq. \eqref{eq:N-body-Lagrangian-GR-1PN} we have dropped divergent quantities involving the self-potential of the  bodies, which can be handled either through Hadamard or dimensional regularization \cite{Blanchet2014}. This procedure can be seen as a \qm{renormalization} of the mass terms \cite{Landau-Lifschitz}. The 1PN-accurate equations of motion stemming from Eq. \eqref{eq:GR-Lagrangian-N-Body} are known in the literature as \emph{Einstein-Infeld-Hoffmann equations} \cite{Einstein1938,Poisson-Will2014}. 

\section{Computation of $2\varepsilon_{ijp} \mathcal{Z}_A^{[ji]}$}
\label{Sec:Appendix-integral-Z}
In this Appendix, we prove that the structure integral $I_p=2\varepsilon_{ijp} \mathcal{Z}_A^{[ji]}/c^2$ (cf. Eqs. \eqref{eq:PPL-spin-equation-1} and \eqref{eq:tensor-mathcal-Z-A-ij}) gives a vanishing contribution to the rotational dynamics. 

We first write $s_{ij}=\varepsilon_{ijp}\xi^p$, where $\xi^p$ is the spin density vector. After performing the resulting computations, we find that $I_p$ is the sum of the following  integrals:
\begin{subequations}
\begin{align} \label{eq:structure-integral-firstterm}
& A_p = \dfrac{6G}{c^2}\varepsilon_{lpk}\int_A \dd^3\boldsymbol{y} \, \dd^3\boldsymbol{y}^\prime \xi^k \xi^{\prime j} \dfrac{(y-y^\prime)^l (y-y^\prime)^j}{\vert \boldsymbol{y}-\boldsymbol{y}^\prime \vert^5}, \\ \label{eq:structure-integral-secondterm}
& B_p = - \dfrac{2G}{c^2} \varepsilon_{lpk}\int_A \dd^3\boldsymbol{y} \, \dd^3\boldsymbol{y}^\prime \xi^k \xi^{\prime l} \dfrac{1}{\vert\boldsymbol{y}-\boldsymbol{y}^\prime\vert^3}.
    \end{align}
\end{subequations}

We consider  the integral \eqref{eq:structure-integral-secondterm} first. By swapping  the  integration variables $\boldsymbol{y}$ and $\boldsymbol{y}^\prime$, it is easy to prove that $B_p=-B_p$ and hence $B_p=0$. Remarkably, from this integral we deduce a crucial property of the spin vectors inside each body $A$, namely 
\begin{equation}\label{eq:parallelism-condition}
    \dfrac{1}{c^2}\int_A \dd^3 \boldsymbol{y}\,\dd^3 \boldsymbol{y}^\prime \boldsymbol{\xi} \times \boldsymbol{\xi}^\prime     = 0.
\end{equation}
In other words, at 1PN order the spin density vectors inside each body point in the same direction.

The integral \eqref{eq:structure-integral-firstterm} requires some additional work. By exploiting the reflection symmetry property, we can realistically describe the body $A$ as a general orthogonal ellipsoid having the axes $a,b,c$ coincident with those of a Cartesian   coordinate system. Applying the following dilation transformation
\begin{align}
(x,y,z)\to (X,Y,Z)=(ax,by,cz),
\end{align}
the orthogonal ellipsoid is mapped into a unit sphere.

Using spherical coordinates and setting the spin vector along the $Z$-axis (cf. Eq. \eqref{eq:parallelism-condition}), we then employ the rotationally invariant property of the sphere. This allows to set, without loss of generality, $\varphi^\prime=\varphi=0$, because the vector $\boldsymbol{y}-\boldsymbol{y}^\prime$ lies always in a plane. In this way, we have $A_X=A_Z=0$, while $A_Y\neq0$, and in fact its expression is (after a coordinate transformation for the angular variables and up to multiplicative constant terms) 
\begin{align} \nonumber
&A_Y = \int \limits_{-\frac{\pi}{2}}^{\frac{\pi}{2}} \dd\theta \sin\theta \int \limits_{-\frac{\pi}{2}}^{\frac{\pi}{2}} \dd\theta^\prime\int \limits_0^1 \dd r \,\int \limits_0^1 \dd r^\prime \sin\theta^\prime g(r,r^\prime,\theta,\theta^\prime),\label{eq:radius}\\ 
&g(r,r^\prime,\theta,\theta^\prime)= \dfrac{r^2 r^{\prime 2} (r\cos\theta - r^\prime\cos\theta^\prime)(r\sin\theta-r^\prime\sin\theta^\prime) }{\left[r^2 + r^{\prime 2} - 2rr^\prime\cos\left(\theta-\theta^\prime\right)\right]^{5/2}}.
\end{align}

The integration with respect to the radial coordinates produces a well-behaved function  in  the domain of integration. Indeed,  $g$ can be dominated by 
\begin{align}  
& \vert g \vert  \le \frac{r}{\sqrt{1-r^2}},    
\end{align}
and
\begin{equation}
\int \limits_0^1 \dd r \,\int \limits_0^1 \dd r^\prime \frac{r}{\sqrt{1-r^2}}=\Biggr{[}-\sqrt{1-r^2}\Biggr{]}^1_0=1.   
\end{equation}
Since Eq. \eqref{eq:radius} will be evaluated in 0 and 1, the final expression will depend only on the  polar angles $\theta$ and $\theta^\prime$. The resulting function is odd on a symmetric domain, entailing thus $I_p=0$.

We stress that the hypothesis of reflection symmetry is extremely important in this computation.  
Indeed,  in the most general case the integral \eqref{eq:structure-integral-firstterm} may be,  in principle, non-vanishing.

\end{document}